\documentclass[a4paper,aps,pra,twocolumn,showpacs,preprintnumbers,amsmath,amssymb]{revtex4}
\usepackage{bbm}
\usepackage{amsmath}
\usepackage{verbatim}
\usepackage{graphicx}

\bibliography{References}
\begin{document}

\title{Generation of a superposition of odd photon number states for
quantum information networks}

\author{J. S. Neergaard-Nielsen$^{(1,3)}$, B. Melholt Nielsen$^{(1,3)}$, C.
Hettich$^{(1,3)}$, K. M\o lmer$^{(2,3)}$, and E. S. Polzik$^{(1,3)}$ }

\affiliation{ $^1$Niels Bohr Institute, Copenhagen University, DK 2100, Denmark \\
$^2$ Department of Physics and Astronomy, University of Aarhus, DK 8000, Denmark \\
$^3$ QUANTOP, Danish Research Foundation Center for Quantum Optics}

\begin{abstract}
We report on the experimental observation of quantum-network-compatible
light described by a non-positive Wigner function. The state is generated
by photon subtraction from a squeezed vacuum state produced by a continuous
wave optical parametric amplifier. Ideally, the state is a coherent
superposition of odd photon number states, closely resembling a
superposition of weak coherent states (a Schr\"odinger cat), with the
leading contribution from a single photon state in the low parametric gain
limit. Light is generated with about 10,000 and more events per second in a
nearly perfect spatial mode with a Fourier-limited frequency bandwidth
which matches well atomic quantum memory requirements. The generated state
of light is an excellent input state for testing quantum memories, quantum
repeaters and linear optics quantum computers.
\end{abstract}

\pacs{03.65.Wj; 03.67.-a; 42.50.Dv}

\maketitle 

Generation of non-classical, non-Gaussian states of light compatible with
atomic quantum memory has been an outstanding challenge driven by various
applications in quantum information processing. Storage of single photon
states in atomic memories is at the heart of proposals for quantum
repeaters \cite{repeater}, whereas Fock states or Schr\"odinger cat states,
stored off-line, form the basis of linear optics quantum computing
\cite{linear1,linear2}. Various approaches to generation of single photon
states compatible with atoms have been pursued: single atoms in free space
\cite{grangierRb} and in high-finesse cavities \cite{Kimble,Rempe}, and
atomic ensembles \cite{kimble04,kuzmich,lukin05}, and non-classical
features such as photon antibunching and violation of classical
inequalities have been demonstrated. However, the dominant contribution to
the detected state of light in all of those experiments has been a vacuum
state, due to either high losses or high mode impurity. High fidelity
single photon states \cite{Lvovsky1,Bellini} have been reported previously
but those were not compatible with atomic memories. Schr\"odinger cat
states have been demonstrated for microwave radiation \cite{Haroche} but
not for light.

In this Letter we report on generation of light, which is compatible with
atomic quantum memories and which is characterized by negative values of
the Wigner function. The generated states are close to cat states with
small coherent amplitude, and converge to a single photon state for certain
experimental parameters. The compatibility with quantum memories is ensured
by the purity of the spatial, temporal modes of the field. The generated
field has a well defined Fourier-limited spectrum with a bandwidth around 9
MHz in the low gain regime, is frequency tunable, and has a perfect spatial
Gaussian profile.

We follow the essence of the proposal \cite{Dakna} in subtracting a photon
from squeezed vacuum by transmitting squeezed light through a weakly
reflecting beam splitter and detecting a photon in the reflected beam. In
\cite{Grangier04} this idea was used for demonstration of generation of
non-Gaussian states. In the absence of losses, the state \cite{Dakna} in
case of, e.g., 3 dB of squeezed input is
$\Psi\sim|1\rangle+0.29|3\rangle+0.09|5\rangle+\ldots $ in the Fock basis.
It has been shown in \cite{Dakna,Lund,Kim} that such a state well
approximates a small-amplitude Schr\"odinger cat state.

\begin{figure}[]
  \begin{center}
    \includegraphics[width=\columnwidth]{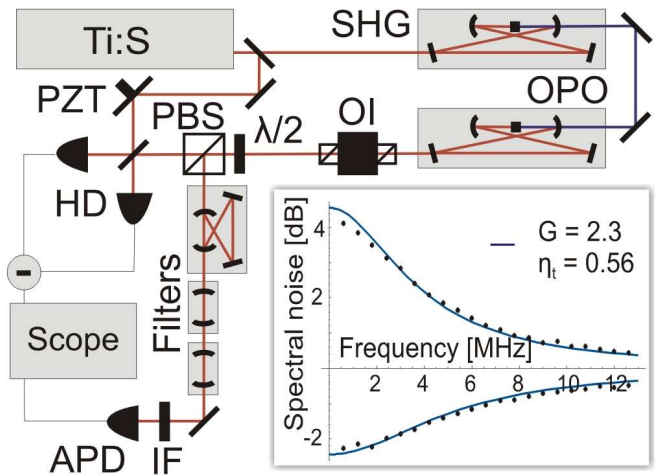}
    \caption{(Color online) Experimental setup. SHG - second harmonic
generator, OI - optical isolator (to prevent scattered LO light from
entering the OPO), $\lambda$/2 - half wave plate, PBS - ``magic'' beam
splitter, PZT - piezo mounted mirror, HD - homodyne detector, IF -
interference filter, APD - avalanche photo diode. The inset shows the
squeezing spectrum obtained from a Fourier analysis of the raw data. The
blue curve is the best fit of the theoretical spectrum. The fitted
parameters (OPO gain $2.3$, and total efficiency,
$\eta_t=0.56$) match well the experimentally estimated values.}
    \label{figure1}
  \end{center}
\end{figure}

In order to generate a narrow-band state of light in a pure spatial mode,
we make use of a continuous wave (cw) optical parametric oscillator (OPO)
below threshold for generation of the initial squeezed field. Such a device
has been successfully used for generation of squeezed and entangled
Gaussian states compatible with atomic targets
\cite{polzik92,georg,hald,schori,schori1}. In an OPO the non-linear optical
medium is placed inside an optical resonator (Fig.\ \ref{figure1}), thus
providing light generated within well defined frequency and spatial modes.
At the same time the resonator significantly enhances the spectral
brightness of the selected modes. All those features prove advantageous for
the OPO-based source compared to single-pass sources \cite{Lvovsky1,
Bellini, Grangier04} as far as compatibility with atomic memories is
concerned. The cw character of the experiment requires a multimode theory
for its adequate description. We have developed this theory based on the
formalism of Gaussian states and Gaussian covariance matrices \cite{Klaus}
(see also an alternative theoretical multi-mode analysis for a different
detection model \cite{Japanese}).

In the experiment (Fig.\ \ref{figure1}) the output of the Ti:Sapphire laser
around 852 nm (close to the Cesium D2 transition) is frequency doubled to
generate up to 150 mW of the blue OPO pump beam (426 nm), while a small
part of the laser beam is split off to act as the local oscillator (LO).
The OPO is an 81 cm long bow-tie cavity with a 10 mm, temperature-tuned
KNbO$_3$ crystal. The down-conversion is Type I with 90$^{\circ}$
non-critical phase matching, hence all down-converted photon pairs are
generated in the same polarization and in the same direction defined by the
spatial mode of the resonator. The OPO cavity is locked on resonance with
the laser frequency, so that the degenerate down-conversion mode is
efficiently generated. The output coupling mirror of the OPO has a
transmission of 12.7\% and the intracavity losses are 1.0\%. In addition to
this, the blue pump induces around 1\% absorption of the IR cavity field
due to the light-induced loss process in the crystal - BLIIRA
\cite{BLIIRA}. Altogether, this gives a cavity bandwidth of 4.4 MHz HWHM
and the output coupling efficiency of the cavity around $\eta_{OPO}=0.86$.
More details of this setup can be found in \cite{polzik92,georg,schori}.

In order to characterize squeezing generated by the OPO and the losses in
the homodyne channel, we measure the squeezing spectrum of the OPO. The OPO
output is mixed on a balanced $50/50$ beamsplitter with a few-milliwatt LO
beam. The differential photocurrent is measured by two low noise
photodetectors and its Fourier transform is digitally obtained. The results
are presented in the inset in Fig.\ \ref{figure1}. Note that squeezing is
observed across the entire theoretical bandwidth of the OPO (with the
exception of frequencies below 30 kHz), and not just within a narrow
frequency band. This is necessary for the photon subtracted state
generation. From this data we infer the OPO gain and the overall losses.
The overall propagation efficiency $0.66$ coincides reasonably well with
$\eta=\eta_{OPO}\eta_{pr}(\eta_{hom})^2$ calculated from the measured
values for $\eta_{OPO}$, propagation efficiency from the OPO to detectors
$\eta_{pr}=0.88$ and homodyne efficiency $\eta_{hom}=0.96$. Our homodyne
detectors have the quantum efficiency of $0.90$ which along with the small
dark current lead to the effective detector efficiency $\eta_d=0.85$,
giving the total efficiency of detection $\eta_{t}=0.60$ (close to
$\eta_{t}=0.56$ inferred from squeezing data).

\begin{figure}[!b]
  \begin{center}
    \includegraphics[width=\columnwidth]{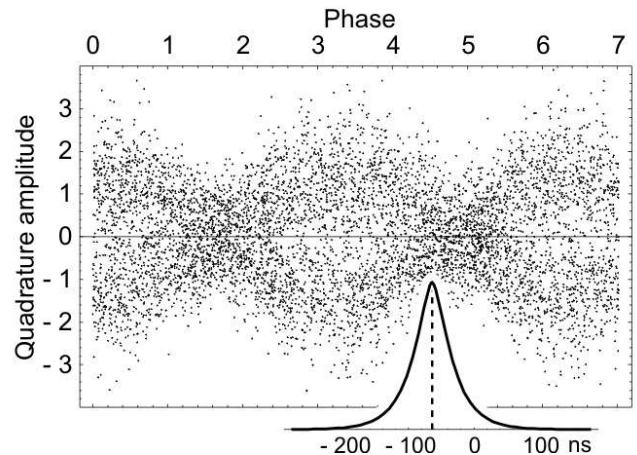}
    \caption{Part of a typical trace obtained after application of the
temporal mode function (shown in the inset) to each of the 20,000 raw data
segments. Each segment of raw data is multiplied by the mode function and
integrated, producing one phase/quadrature point for the trace. The phase
of the LO has been scanned at roughly 1$\pi$/s. The temporal mode function
is shifted with respect to the APD click due to the delay in electronics. }
    \label{figure2}
  \end{center}
\end{figure}

\begin{figure*}[t]
  \begin{center}
    \includegraphics[width=\textwidth]{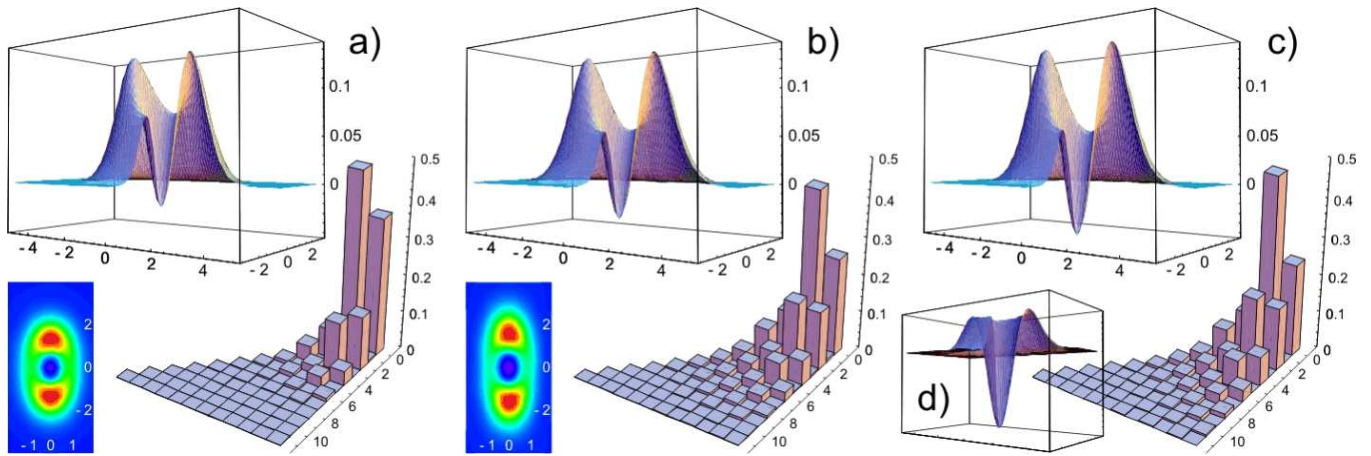}
    \caption{(Color online) Figures a,b) show the density matrices and
Wigner functions (including a top-down contour plot) reconstructed from
experimental data. a) the OPO gain $1.8$ b) gain $2.3$. The presented plots
are averages of 50 separate states, each reconstructed from a 20,000 points
quadrature trace as explained in the text and corrected for the 85\%
detector efficiency. c) state calculated from full multi-mode theory with
the same temporal mode function and the same overall propagation efficiency
$\eta=0.64$ as for the experimental state b). d) state calculated with
perfect efficiency $\eta=1$.}
    \label{figure3}
  \end{center}
\end{figure*}

For the photon subtraction experiment the down-converted output of the OPO
is directed onto a ``magic'' beam splitter (polarizing beam splitter cube
in combination with a half-wave plate) which reflects a small portion
(3-7\%) of the light towards a photon counter - APD (Perkin-Elmer
SPCM-AQR-13, dark count rate 160 s$^{-1}$). Note that for low reflectivity
of the ``magic'' beam splitter the probability of a two-photon click is
very low, i.e., a click of the APD almost for sure corresponds to the
single photon detection. Besides the degenerate squeezed mode the OPO
generates thousands of, non-degenerate, modes which are also phase matched
and resonant with the cavity. Photons in those modes produce unwanted
counts in the photon subtraction channel, and therefore have to be filtered
out. Towards this end, on the way to the APD, the light is filtered with
three consecutive cavities and a spectral filter which block all the OPO
output modes except for the degenerate squeezed mode (Fig.\ \ref{figure1}).
The OPO cavity and the three filter cavities are tuned in resonance with
the degenerate OPO mode by injecting a weak coherent seed beam into the
OPO. The count rate of the APD is then used to produce the error signal for
the frequency locking. A LabView program monitors this count rate while it
scans all cavities at different low frequency rates until the program finds
a setting for the four piezo control voltages which gives an expected
maximal count rate. Under stable lab conditions the lock is reliable down
to a count rate of approximately 5000 s$^{-1}$. With the cavities lock
points established, we inject the blue pump into the OPO and turn down the
coherent seed beam keeping the lock operational now with only the
down-converted light. By turning the ``magic'' beamsplitter to full
reflection, we measure the maximum count rate as a reference for the
subsequent setting of the beamsplitter at the chosen reflection ratio. At
the same time we get an estimate for the total production rate of photons
into the degenerate mode (around $2*10^6\ \mathrm{s}^{-1}$ for the
parametric gain $2.3$).

The LO phase is scanned to perform the tomography of the state. For every
click of the APD, a data segment of 1000 points is recorded within a period
starting 1 $\mu$s before and ending 1 $\mu$s after the trigger event.
20,000 such segments are stored and comprise one full measurement of the
different quadratures of the quantum state. The total acquisition time of a
full measurement is only around 3 seconds and a new measurement can be
initiated a few seconds later. The high spectral brightness of our source
thus allows faster data acquisition than in single pass experiments where
it may take hours.

During the initial data processing each 2 $\mu$s segment of experimental
data is turned into a single quadrature measurement using a particular
temporal mode function.  If the OPO is in the intermediate to high gain
regime, the field populates many modes, and the optimum mode function
should be chosen as the one populated by a quantum state with the strongest
non-classical features, e.g., the one with the largest negative value
attained by the Wigner function. A fairly simple Ansatz for the mode
function $\kappa\, e^{-2\pi\gamma|t|}-\gamma\, e^{-2\pi\kappa|t|}$ , which
is the Fourier transform of the product of two Lorentzians with half-widths
$2\pi\gamma$ and $2\pi\kappa$, shows close to the optimal performance. The
optimal $\gamma$ and $\kappa$ are close to $2\pi *9$ MHz (OPO width) and
$2\pi *48$ MHz (filter cavity width). The temporal shape of the generated
state is shown in Fig.\ \ref{figure2}. Every 2 $\mu$s/1000 points data
segment is then multiplied by the mode function. This gives a number which
is the measured quadrature amplitude value for that segment. This procedure
is equivalent to the use of a pulsed local oscillator with the temporal
shape of the mode function. A full measurement set consists of 20,000
quadrature values, along with their corresponding phase values (obtained
from the time of their acquisition via an independent phase calibration). A
data set obtained in this way is presented in Fig. \ref{figure2}. The lack
of points around zero values of the quadrature phase amplitudes is a clear
signature of the non-Gaussian character of the radiation. We perform the
same procedure for the vacuum state, and use the result for normalization.

To reconstruct the density matrix from the data we use the iterative
maximum-likelihood method of \cite{Lvovsky}, based on \cite{MaxLik} - for
reviews of this and other methods, see \cite{TomographyReviews}. In order
to obtain the Wigner function of the optical field irrespectively of the
detector inefficiency we correct the density matrix for the detector
efficiency using the algorithm of \cite{Lvovsky}. The density matrices and
the Wigner functions of the photon subtracted squeezed vacuum states for
two different degrees of squeezing are shown in Fig.\ \ref{figure3}a,b.
Each presented state is an average of 50 individually acquired 20,000-point
measurements. The averaged values of the Wigner functions at the origin
are, respectively, $W(0,0)=-0.028\pm0.010$ and $W(0,0)=-0.040\pm0.014$ (we
note that even the states reconstructed without correction for detector
efficiency have small negative dips). The negative regions of the Wigner
functions show their profoundly non-classical character. The Wigner
functions are elongated along the direction of the parent squeezed states
signifying the presence of three photon and five photon contributions. In
the limit of very low degree of squeezing the elongated shape would be
reduced to a round shape characteristic for the single photon state. Fig.\
\ref{figure3}c shows the state calculated following our theoretical model
with the efficiency $\eta=0.64$ corresponding to the experimental
efficiency of the state shown in Fig.\ \ref{figure3}b. This number is a
product of $\eta=0.66$ obtained from the squeezing data and an extra factor
$0.97$ which partially accounts for dark counts of the APD. As seen from
Fig.\ \ref{figure3}b and \ref{figure3}c, theory and experiment are in good
agreement. The small discrepancy may be attributed to an admixture of other
modes in the trigger channel.

Due to the cw character of the squeezing produced in the experiment, the single mode theory of
\cite{Dakna,Lund,Kim,Jeong} is not fully applicable. We have developed an exact theoretical treatment,
which is possible because the output from the OPO is a Gaussian state fully described by its second
order moments, i.e., by the two-time correlation functions of the output $\langle
\hat{b}^\dagger(t)\hat{b}(t')\rangle$ and $\langle \hat{b}(t)\hat{b}(t')\rangle$ \cite{Gardiner}. If
the transmission (reflection) of the ``magic'' beam splitter is $|\tau|^2$ ($|\rho|^2=1-|\tau|^2$) and
if the filters for the trigger field can be modeled by a single Lorentzian of half width $2\pi
\kappa$, the annihilation operator for the mode, causing the click of the APD at $t=t_c$ can be
expressed as $\hat{a}_{t_c}=2\rho\sqrt{\pi\kappa}\int_{-\infty}^{t_c} \exp(-2\pi\kappa(t_c-t))
\hat{b}(t) dt + \hat{F}_{\textrm{noise}}$ where $\hat{F}_{\textrm{noise}}$ represents vacuum field
contributions from the ``magic'' beam splitter and from the frequency filters, and its precise form
does not need to be specified. For an arbitrary mode function $u(t)$, we obtain a similar expression
for the annihilation operator of the field mode extracted by the corresponding LO detection within the
time mode $u(t)$, $\hat{a}_{u}=\tau \int_{-\infty}^{\infty} u(t) \hat{b}(t) dt +
\hat{G}_{\textrm{noise}}$, where $\hat{G}_{\textrm{noise}}$ is a vacuum noise term. Losses and
detector efficiencies are incorporated by reduction factors on the first terms and additional vacuum
noise contributions in the equations for $\hat{a}_{t_c}$ and $\hat{a}_{u}$. The second moments of the
four field quadrature variables for these operators are directly obtained from the output correlation
functions and the filter and mode functions. Defining the $4\times 4$ covariance matrix $\gamma$ by
its elements $\gamma_{ij}=2 \textrm{Re}\langle y_i y_j\rangle$, with the four Hermitian field
quadratures forming a column vector ${\bf y}=(y_1,y_2,y_3,y_4)^T$, the Wigner function can be written
explicitly, $W({\bf y})=1/(\pi \sqrt{\textbf{det}(\gamma)})\exp(-{\bf{y}}^T\gamma^{-1}\bf{y})$. This
Gaussian Wigner function fully characterizes the quantum state of the trigger and LO modes prior to
the APD detection event. The action of a photo detection event is described by application of the
corresponding field annihilation operator to the quantum state, which in the Wigner function
representation is represented by a combination of differentiation and multiplication by the $y_i$
arguments \cite{Gardiner}. The reduced state associated with the homodyned field mode is subsequently
obtained by tracing over the trigger field state, i.e., by integrating over the trigger mode
quadrature arguments in $W$. These operations can be done explicitly, and the resulting single mode
phase space function is readily obtained (Fig.\ \ref{figure3}c). For further discussions of this
treatment, see \cite{Klaus}.

The state generated by photon subtraction from single mode squeezed vacuum
is a superposition of odd-number Fock states \cite{Dakna}, which overlaps
with a cat state $|\alpha\rangle-|-\alpha\rangle$ with better than $99\%$
fidelity for coherent amplitudes less than $\alpha=1.2$. In the present
\textit{cw} experiment we have several modes populated, hence we are bound
to get contributions from even number states, and e.g. the vacuum, which
grow with the degree of squeezing. The states shown in Fig.\
\ref{figure3}a,b overlap with cat states with coherent amplitudes
$\alpha=1.05$ and $\alpha=1.3$ with a fidelity of $53\%$. For our
squeezing, the overlap between the lossless state (Fig.\ \ref{figure3}c)
and a cat state would be still quite good, $0.95$ and $0.90$, respectively.
A major improvement of the purity of cat states can be achieved by reducing
optical losses. It seems feasible to obtain an OPO escape efficiency
$\eta_{OPO}=0.95$, a propagation efficiency $\eta_{pr}=0.97$ and a homodyne
efficiency $\eta_{hom}=0.99$, thus reaching a value close to 0.87 for the
total efficiency. Even higher purity can be obtained with purification
procedures proposed in \cite{Lund,Jeong}.

In conclusion we have demonstrated experimentally a new type of
non-classical state of light, the photon subtracted squeezed vacuum state
based on an optical parametric oscillator below threshold. The results are
in good agreement with a complete multi-mode theoretical analysis. With the
time mode function of 100 nsec, and the transform limited bandwidth around
10 MHz, frequency tunable around Cesium D2 transition, the source is
compatible with atomic memories and networks.

After submission of this paper, generation of an optical ``kitten'' state
was reported by Ourjoumtsev et al. \cite{Grangier06}.

This research has been funded by EU grants COVAQIAL and QAP.

\end{document}